# Competing dynamics of single phosphorus dopant in graphene with electron irradiation

Cong Su[1,3,*], Mukesh Tripathi[2], Qing-Bo Yan[4], Zegao Wang[5], Zihan Zhang[6], Leonardo Basile[7], Gang Su[6], Mingdong Dong[5], Jani Kotakoski[2], Jing Kong[3,8], Juan-Carlos Idrobo[9], Toma Susi[2,*], and Ju Li[1,*]

[1] *Department of Nuclear and Materials Science and Engineering, Massachusetts Institute of Technology, Cambridge MA 02139, USA*

[2] *University of Vienna, Faculty of Physics, Vienna 1090, Austria*

[3] *Research Lab of Electronics (RLE), Massachusetts Institutes of Technology, Cambridge MA 02139, USA*

[4] *College of Materials Science and Opto-Electronic Technology, University of Chinese Academy of Sciences, Beijing 100049, China*

[5] *Interdisciplinary Nanoscience Center (iNano), Aarhus University, Aarhus 8000, Denmark*

[6] *School of Physical Sciences, University of Chinese Academy of Sciences, Beijing 100049, China*

[7] *Department of Physics, Escuela Politécnica Nacional, Quito 170517, Ecuador*

[8] *Department of Electrical Engineering and Computer Science, Massachusetts Institute of Technology, Cambridge MA 02139, USA*

[9] *Center for Nanophase Materials Sciences, Oak Ridge National Laboratory, Oak Ridge TN 37831, USA*

[*]Corresponding authors: liju@mit.edu (J.L.), toma.susi@univie.ac.at (T.S.), csu@mit.edu (C.S.)

**Abstract**

Atomic-level structural changes in materials are important but challenging to study. Here, we demonstrate the dynamics and the possibility of manipulating a phosphorus dopant atom in graphene using scanning transmission electron microscopy (STEM). The mechanisms of various processes are explored and compared with those of other dopant species by first-principles calculations. This work paves the way for designing a more precise and optimized protocol for atomic engineering.

Imperfections and their dynamics [1,2] have a profound effect on many materials properties. Such defects can be either intrinsic or purposefully engineered, and control over them on the atomic level is the ultimate limit of materials science. Historically, scanning tunneling microscopy (STM) has been used to manipulate adatoms on the cryogenic surface of metals [3], and later, atomic force microscopy was used in a similar manner, but at room temperature [4,5]. However, mechanical manipulation is inherently slow and hard to harness for real scalable applications. Recently, following an understanding of their dynamics [6], first Susi et al. and then Dyck et al. demonstrated that Si dopants are controllable in graphene using focused electron beams in the context of STEM [7,8]. As the graphene-dopant system is stable under room temperature, this technique potentially emerges as a fundamentally new tool for atomic engineering, with a performance already nearly on par with STM [9]. Additionally, atom-by-atom defect creation and manipulation might also create extended functional graphene structures that are hard to chemically synthesize [10,11]. However, despite the importance of dynamics of graphene dopants under electron irradiation, only limited data has been reported up to now [12–15].

Doping with phosphorus (P) has been suggested as a means to modify the electronic properties of carbon nanotubes [16,17] and graphene [18,19]. P is thought to be effective in catalyzing oxygen reduction [20,21], and is expected to be a sensitive detector of toxic gas [22]. Due to its non-zero nuclear spin, P might also be useful in quantum informatics, similar to the use of the nitrogen vacancy (NV) center as a small nuclear magnetic resonance detector [23]. An investigation of the configurations and dynamics of P dopants is required for realizing such applications. Aberration-corrected STEM provides a powerful way of identifying dopant atoms with atomic-resolution imaging and electron energy-loss spectroscopy (EELS) [24,25]. Single Si [26,27], N [27,28] and B [29,30] dopant atoms in graphene have already been imaged and their chemical bonding distinguished from the EELS near-edge fine structure. Substitutional P atoms introduced by low-energy ion implantation have also been directly observed in graphene, but the initial samples suffered from severe contamination on the surface and topological lattice defects created by the ions [31]. Thus, a viable way for fabrication and large scale production of devices from tailored graphene structures is still needed.

In this work, we use STEM to image, identify and manipulate P impurities in single-layer graphene synthesized via chemical vapor deposition (CVD), which is found to be largely free of contamination. Four types of dynamical processes are observed: *direct exchange* and *Stone-Wales (SW) transitions* modify the lattice structure; *knock-out* of a C neighbor and *replacement* of P by C modify the chemical stoichiometry. Our simulations show that the P atom will not move unless there is an out-of plane momentum transfer from an incoming electron to a C atom neighboring the dopant. Although the calculations indicate that the lowest energy dynamical process is a SW transition (followed by direct exchange and then a knock-out), the experimental observations of the SW transition are rare. The theory shows that direct exchange is more commonly observed because the direction of the momentum required for this process is mostly normal-to-plane, which is also the direction of the imaging electrons. It is also shown that an additional layer of graphene and a slight tilt of the sample (5-10°) could result in an improved control of P impurities in graphene.

A detailed description of sample fabrication and characterization methods can be found in the Supplementary Materials (SM). The P-doped graphene is transferred onto Quantifoil Au TEM grid using the direct transfer method [32] and then characterized by a Nion UltraSTEM 100 equipped with a cold field emission gun operated at 60 kV. All dopants are further identified with EELS. *Ab-initio* molecular dynamics (ab-MD) is performed using density functional theory (DFT) within the general gradient approximation (GGA), in the form of Perdew-Burke-Ernzerhof's exchange-correlation functional [33]. A 1 fs time step used in the ab-MD simulation was tested to produce trajectories very similar to those obtained with shorter ones. All simulations were performed with Vienna Ab-initio Simulation Package (VASP) [34]. The Atomeye visualizer was used during calculation [35] and VESTA for rendering graphics [36].

In Fig. 1, the four types of dynamic processes are shown, categorized into two groups: 1) atom-conserved *hopping*: including *direct exchange* (Fig. 1(a), earlier dubbed "bond inversion" in the context of Si [6]) and *SW transition* (Fig. 1(b)) of the two neighboring P and C atoms [37]; 2) atom-non-conserved *hopping termination*: including *knockout* (Fig. 1(c)) where one of the neighbor C atoms of P is knocked out by the electron beam and disappear from the scene, and *replacement* (Fig. 1(d)) where P is replaced by a C adatom, apparently aided by the electron beam. After these hopping termination steps, the P atom is experimentally found to no longer able to jump (under 60 keV electron beam). By performing ab-MD and climbing-image nudged elastic band (cNEB) simulations [38], we provide an explanation for the mechanism of these electron-beam-aided dynamic processes, and supply guidance for the manipulation of P.

In Fig. 1(a), three consecutive frames of direct exchange including a transition frame are recorded. As a result, the P dopant atom exchanges site with its neighbor C atom while the electron beam is scanning right across the C (the white dashed line). In Fig. 1(b), the SW transition is accompanied by a direct exchange at the start. After the direct exchange (frame 1 to 2), the P-C bond is rotated by 90° (frame 2 to 3), and the original honeycomb lattice distorts into the transition state with two pairs of 5- and 7-membered rings (55-77 structure hereafter). The 55-77 structure is only stable for about 0.2 s before reverting to the original state (frame 3 to 4) due to continuing electron irradiation. It should be noted that for reasons explained below, the SW transition of P is rarely seen in experiment. In Fig. 1(c), the three-coordinated P (frame 1) turns into four-coordinated (frame 2) when a neighboring C atom is knocked out by the electron beam. Once this happens, the P atom becomes immobile. In Fig. 1(d), P is replaced by C, which is the commonly observed fate of P impurities under intense electron irradiation—in stark contrast to Si, which are almost never removed. It should be noted that we never observed P knocked out leaving a vacancy behind, as expected since its displacement cross section as a heavier atom is several orders of magnitude smaller than the C atoms.

To explain how these processes are initiated, we performed ab-MD and cNEB calculations. Figs. 2(a)-(d) are four examples representing different dynamical processes shown in the order of: unchanged, knock-out, direct exchange, and SW transition. It is found that all of these dynamics of P dopants are initiated by an out-of-plane momentum of a C neighbor, similar to what is shown by Susi et al. for Si [6]. To clarify the spherical coordinates we use here, definitions of $\theta$ and $\varphi$ are plotted in the first frame of Fig. 2(a), along with an example of unchanged structure ($\theta = 20°$, $\varphi = 45°$, with the kinetic energy of C atom $E_C=15.0$ eV), which comes back to the original configuration after the dynamic process. As an example of knock-out in Fig. 2(b), the initial momentum of C atom is tilted toward $\theta = 20°$, $\varphi = 180°$, with $E_C$ increases to 17.0 eV. In Fig. 2(c), an initial velocity perpendicular to the plane ($\theta = 0°$) and $E_C = 17$ eV cause a direct exchange. If the initial velocity is not strictly upwards, but tilted at an angle ($\theta = 15°$, $\varphi = 135°$, $E_C$=16 eV as in the example), SW transition happens (Fig. 2(d)) [37]. To find out the distribution of different processes as a function of initial angles and kinetic energies of the C atom, we performed ab-MD calculations at $E_C$=15, 16, and 17 eV. For each energy, the angular space is sampled with an interval of 15° for the azimuthal angle $\varphi$ and 5° for the polar angle $\theta$ (up to 25°), and results plotted in Fig. 2(e)-(g).

Several conclusions can be made from these plots: (I) a SW transition can be initiated with a lower energy (starting from 15 eV) than direct exchange. (II) As $E_C$ increases, direct exchange gradually becomes the dominant dynamical process. (III) When $E_C$ reaches around 17 eV, knock-outs begin to occur. (IV) Somewhat counterintuitively, direct exchange is easier when the initial momentum is pointing away from ($\varphi = 180°$), instead of pointing toward ($\varphi = 0°$) the target C atom.

In the experiments, we found P to hop much less actively than what has been reported for Si [9]. To explain this, we compare the energy range of direct exchange for Si, P, as well as Al when assuming a head-on collision ($\theta = 0°$; Fig. 3(a)). Even though the full energy range is not limited to $\theta = 0°$ being shown, it serves as a point of comparison between the different elements. In this comparison, Si clearly covers the greatest energy range for direct exchange; as a result, its probability is much larger for Si than for P. The displacement threshold of C neighbor of Al dopant is much lower than the rest of two, so knock-out is a more likely event. In fact, we have observed Al dopant and its surrounding atoms to be displaced by 60 keV electron beam (Fig. S5), while we never observe such process for Si or P. This suggests that a lower acceleration voltage could help to facilitate direct exchange also for Al.

On the contrary, a SW transition is more likely to happen for a P dopant, while it is never observed for Si dopants. cNEB calculations that explain this are shown in Fig. 3(b). As a broader comparison, we compute 6 elements, both of which could initiate SW transition. To be able to observe the SW process, the 55-77 structure should be sufficiently stable under the electron beam. Its stability is proportional to the energy barrier between the highest energy transition state and the 55-77 structure, which is shown as the activation energy $E_a$ (the original cNEB curves can be found in Fig. S6 in SM). The stability of 55-77 structures follows the order C>N>B>P>Si>Al. According to Arrhenius theory, the transition rate from the 55-77 structure to original honeycomb lattice is 14 orders of magnitude higher for Si than P due to 0.8 eV of barrier difference; hence, the 55-77 structure of Si is almost impossible to capture in experiments.

As a common ending point of P dynamics in STEM, replacement by C plays an important role. It is widely accepted that free C adatoms travel on the surfaces of graphene under electron microscopy conditions [6,39]. In Fig. 3(c), our calculation shows that C adatoms can bond stably on a C-C bridge close to the underside of a P site (shown as the initial state). By performing a cNEB calculation, we see that to transit from this initial state to a final state where the P has been replaced by C, the system only needs to cross a 0.4 eV barrier, easily available from the 60 keV electron beam [40], subsequently reducing the total energy of the system by 4.5 eV. Further, the initial configuration is 0.33 eV lower in energy than an adatom bound directly on top of the impurity—in contrast to Si where the top site is 0.23 eV lower in energy, possibly explaining why we never observe Si being replaced by C. More details can be seen in the Fig. S7 in SM.

It is interesting to note that a P atom is much harder to be replaced in a double-layer configuration (Fig S7 in the SM), where atom diffusion on one side is suppressed. During the course of our STEM observation (at least 12 min continuously with an irradiation dose rate comparable to our monolayer data), the P dopant in a double-layer was not replaced by C atom, whereas in a monolayer it typically survives less than 3 min. We believe that this is because the second layer prevents free mobile C atoms from reaching the opposite side of the P dopant; therefore, the replacement process is suppressed.

To intentionally manipulate the P dopant, we tried to initialize the direct exchange by targeting the electron beam at a neighbor C atom. The initial position of P dopant is shown in Fig. 4(a). The yellow cross indicates where the electron beam is parked for 10 s, and afterwards, a second frame is captured immediately, shown in Fig. 4(b). As a result, the P atom hops site as expected, but this occurred only after 12 unsuccessful iterations. Comparing with Si impurities, P is much harder to move: parking the electron beam on the neighbor C site instead typically triggers the replacement process. In total, we tried to manipulate nine P impurities, only one of which jumped, one lost a C neighbor, and seven were replaced by C after on average 22±5 (mean±std. err.) 10-second spot irradiations.

The angular distribution of direct exchange and the long lifetime of a P dopant in a double-layer suggests that the it may be better controlled with the addition of a second graphene layer, and by tilting the sample such that the electron beam has an angle with respect to the graphene surface of 5–10°. A better strategy for inducing direct exchange might be designed from the scheme we propose in sections 8 and 9 of the SM (including reference [41]).

In summary, we have observed four types of dynamics of P dopants in graphene, and explain the mechanisms for each process by first-principles calculations, providing a convenient categorization of dynamics of other impurity atoms as well. We have also demonstrated the possibility of electron-beam manipulation of a P dopant, albeit significantly more challenging than for Si. The analysis presented here can further help developing techniques for controlling P and other dopants in graphene at room temperature and with atomic precision using scanning transmission electron microscopy.

**Acknowledgement**
J.L. and C.S. acknowledge support by NSF DMR-1410636 and ECCS-1610806. J.K. (MIT) and C.S. acknowledge the support from U.S. Army Research Office through the MIT Institute for Soldier Nanotechnologies (Grant No. 023674). T.S. and M.T. acknowledge funding by the Austrian Science Fund (FWF) via project P 28322-N36, and T.S. further by the European Research Council (ERC) under the European Union's Horizon 2020 research and innovation programme (grant agreement No. 756277-ATMEN). J.K. (Univ. of Vienna) was supported by the FWF via project I3181-N36, and by the Wiener Wissenschafts-, Forschungs- und Technologiefonds (WWTF) project MA14-009. The electron microscopy experiments were conducted at the Center for Nanophase Materials Sciences, which is a DOE Office of Science User Facility (J.C.I.). M.D. acknowledge grants from the Danish Council for Independent Research, AUFF NOVA project from Aarhus Universitets Forskningsfond and EU H2020 RISE 2016-MNR4SCell project. Q.B.Y. and G.S. are supported in part by the MOST of China (Grant No. 2013CB933401), the NSFC (Grant No.14474279), and the Strategic Priority Research Program of the Chinese Academy of Sciences (Grant Nos. XDB07010100, XDPB08). L.B acknowledges support by Escuela Politécnica Nacional (EPN) via project PIJ-15-09.

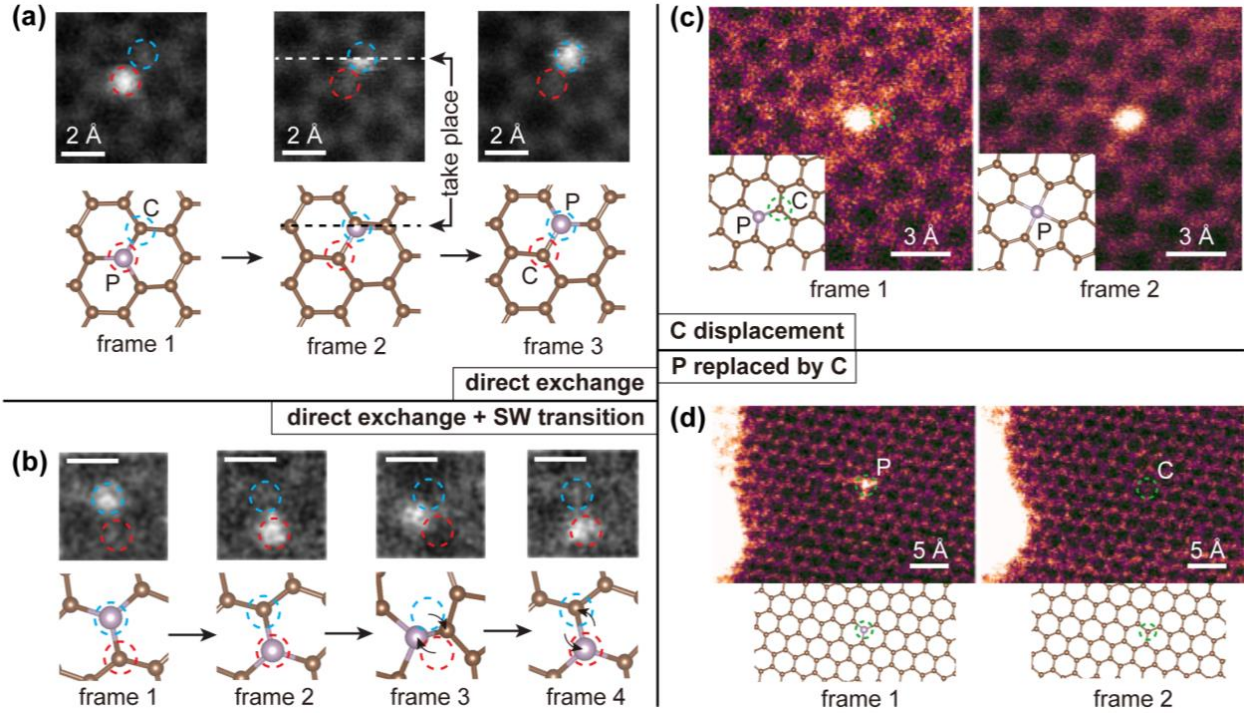

FIG. 1. Competing P dopant dynamics in graphene lattice. (a) Three frames showing direct exchange with the initial (frame 1), transition (frame 2), and final configurations (frame 3). White and black dashed lines indicate the row of the beam scanning when the exchange event happens. Scan speed: 8.4 s/frame. (b) Four frames showing both direct exchange (frame 1 to 2) and Stone-Wales (SW) transition (frame 2 to 4). Scan speed: 0.07 s/frame. (c) Neighboring C atom knocked out by electron beam, turning the three-coordinated P into four-coordinated. Scan speed: 8 s/frame. (d) P dopant being replaced by a C atom. Scan speed: 4 s/frame. The different image color-codings represent different categories: grey represents atom-conserved processes, and magenta represents atom-non-conserved processes. Blue and red dashed circles in (a) and (b) represent non-equivalent lattice sites of graphene, and the green dashed circles in (c) and (d) indicate the location of non-conserving process. Scale bars in (b): 2 Å.

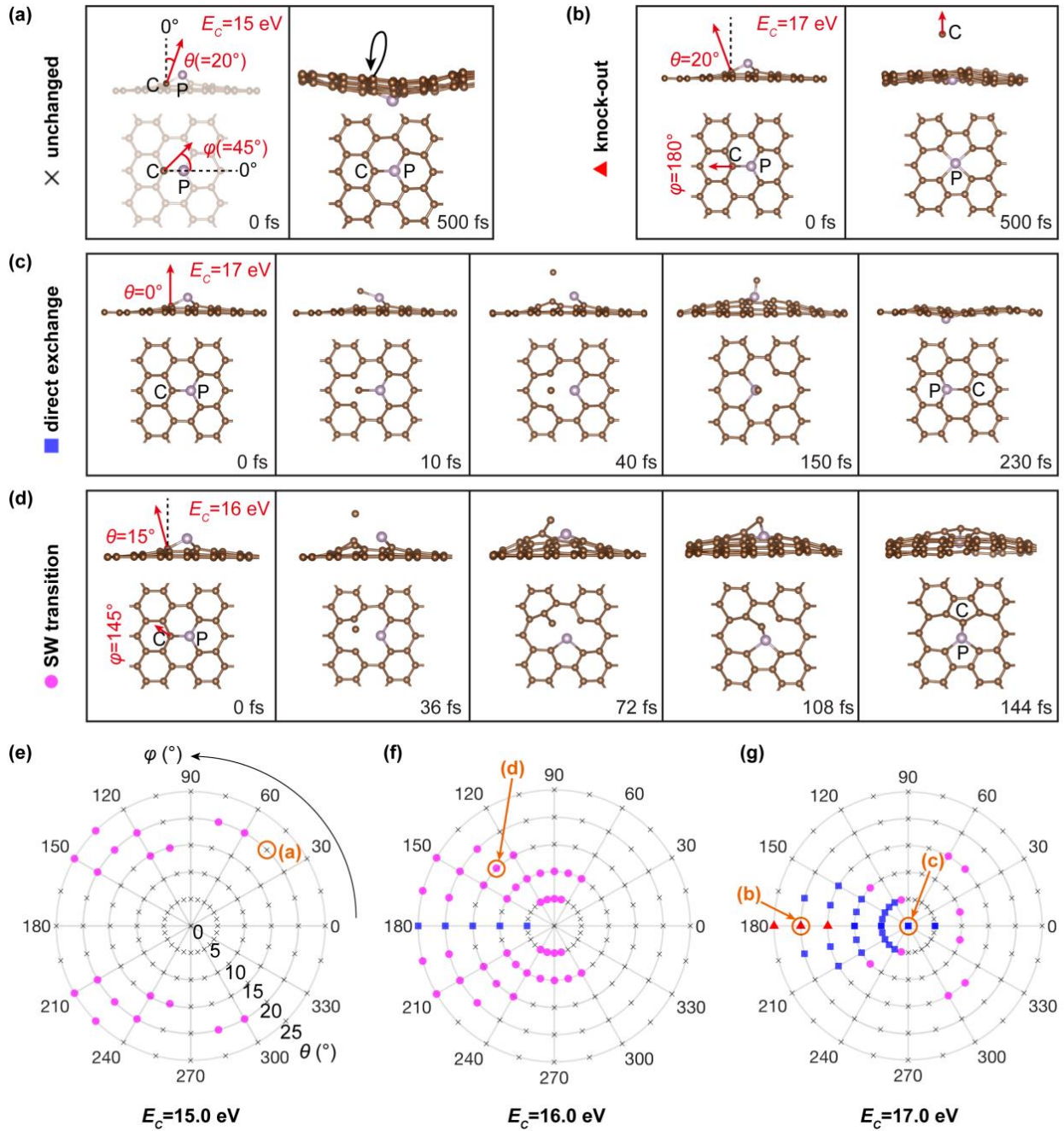

FIG 2. Snapshots of molecular dynamics simulations when an out-of-plane momentum has been injected to a neighbor C atom to the P impurity resulting in an (a) unchanged structure ($\theta = 20°$, $\varphi = 45°$, $E_C$=15.0 eV), (b) knock-out ($\theta = 20°$, $\varphi = 180°$, $E_C$=17.0 eV), (c) direct exchange ($\theta = 0°$, $E_C$=17.0 eV) and (d) SW transition ($\theta = 15°$, $\varphi = 135°$, $E_C$=16.0 eV). The red arrows indicate the direction of projected momentum of the C atom along the in-plane and normal-to-plane directions (lengths not to scale), with the definition of spherical coordinate angles $\theta$ and $\varphi$ shown

in the first frame of (a). (e-g) Angular distribution maps of different possible lattice transitions obtained when a neighbor C atom to the P impurity is injected with an initial out-of-plane momentum corresponding to a kinetic energy $E_C$ of (e) 15.0 eV, (f) 16.0 eV, and (g) 17.0 eV. The symbols indicate the angular components of the initial momentum that produce either knock-out (red triangles), direct exchange (blue squares), SW transitions (magenta circles) or an unchanged lattice (black crosses). The orange circles in (e-g) mark the data points representing examples in (a-d).

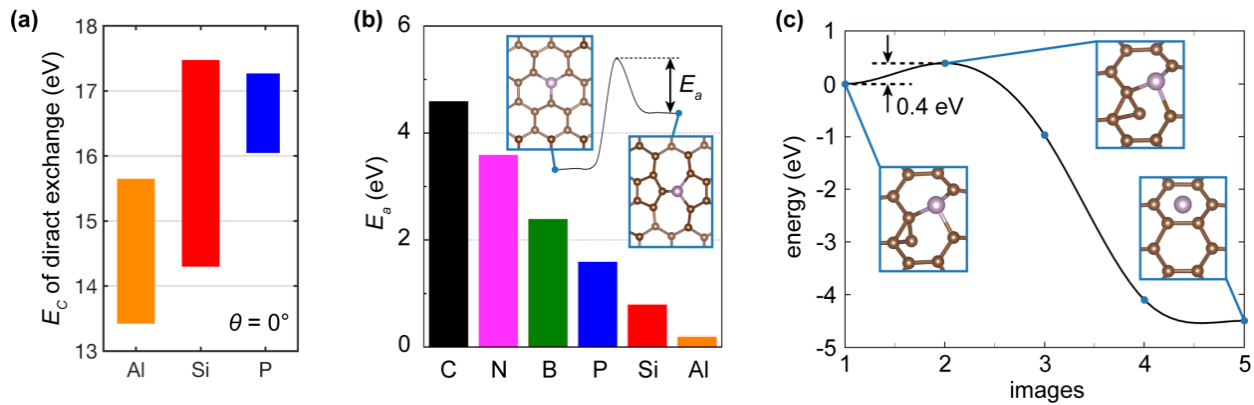

FIG. 3. (a) Comparison of the direct exchange energy ranges of the elements Al, Si, and P, when an upward knock-on happens ($\theta = 0°$). (b) The energy barriers ($E_a$) of configurational change from 55-77 structures back to the honeycomb for various elements are illustrated (C: 4.6 eV, N: 3.6 eV, B: 2.4 eV, P: 1.6 eV, Si: 0.8 eV, Al: 0.2 eV). Inset: The definition of $E_a$ indicated in the energy profile of SW transition, where the left side indicates the honeycomb lattices and the right side is the 55-77 structure. The details of cNEB curves can be found in Fig. S6 of the SM. (c) cNEB barrier for the proposed mechanism of P dopant replacement by C. Insets: the initial, saddle-point, and final configurations.

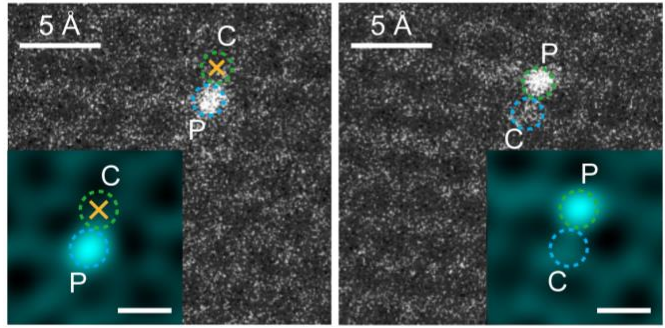

FIG. 4. Intentional manipulation of a P dopant. The yellow crosses indicate the location where the electron beam was parked for 10 s to purposefully move the P atom by one lattice site. Green and blue dashed circles indicate the two non-equivalent lattices sites of graphene. Insets: the region of interest after applying a Gaussian filter. Scale bars are 2 Å.

Supplementary Materials

# Competing dynamics of single phosphorus dopant in graphene with electron irradiation

## 1. Methods

*Sample fabrication*. P-doped graphene is synthesized using chemical vapor deposition. Firstly, a 25 μm-thick Cu foil (Alfa Aesar, no. 13382) was washed in 5% HCl solution for 3 min, and then rinsed in DI water several times. After that, the Cu foil was dried by nitrogen and quickly loaded into a tube reactor (1 inch diameter, 1.5 m length). A quartz boat container with about 100 mg of triphenylphosphine ($C_{18}H_{15}P$, Sigma Aldrich) used as a sole precursor source was placed upstream from the sample as shown in Figure 1. The system was evacuated to a vacuum lower than $1\times10^{-3}$ Pa. The zone-1 of the furnace was first heated to 1050 °C at a rate of 20 °C/min in 25 sccm $H_2$ and 100 sccm Ar. After annealing for 20 min, the temperature was decreased to 1000 °C. Then, zone-2 of the furnace was heated to 80 °C at a rate of 5 °C/min. The triphenylphosphine vapor is carried into zone-1 by the flowing $H_2$ and Ar, initiating graphene growth on the Cu foil. After 20 min, the system is cooled to room temperature with a cooling rate of about 50 °C/min by opening the furnace. During growth and cooling, the flux of $H_2$ and Ar remains unchanged. The P-doped graphene is then transferred onto Quantifoil Au TEM grids using the direct transfer method for electron microscope imaging.

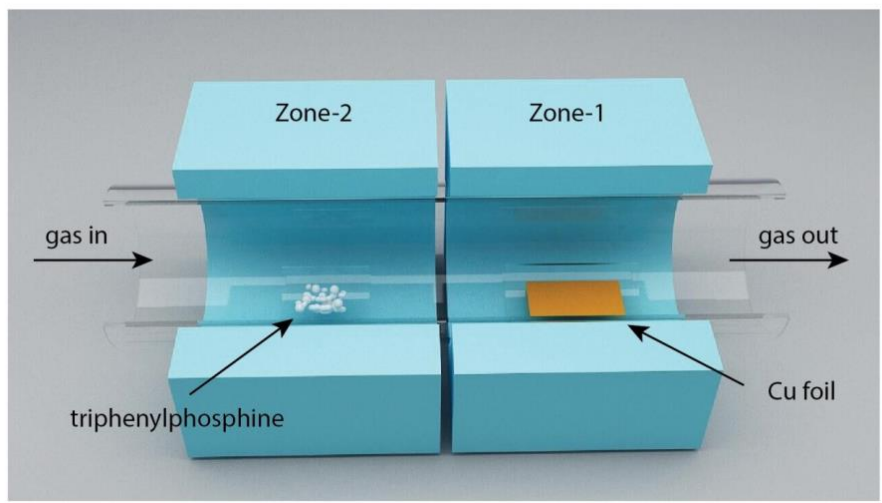

**Fig. S1**. A schematic of the P-doped graphene synthesis reactor.

*Raman Characterization.* Firstly, PMMA was spin-coated on its surface and the Cu foil etched by FeCl$_3$ solution. After washing in DI water several times, the PMMA/graphene was transferred onto a SiO$_2$/Si substrate, and baked at 180 °C for 5 min. The PMMA was then removed in warm acetone. A typical Raman spectrum is shown in Figure S2, collected using a Renishaw Raman spectroscope (laser excitation 514 nm). Both pristine graphene and P-doped graphene show two intense Raman features, which are assigned to G (~1585 cm$^{-1}$) and 2D (~2685 cm$^{-1}$) peaks. Significantly, P-doped graphene presents also a strong D (~1348 cm$^{-1}$) and D' (~1620 cm$^{-1}$) peaks, which are activated by defects such as in-plane heteroatom substitutions, vacancies, or grain boundaries.

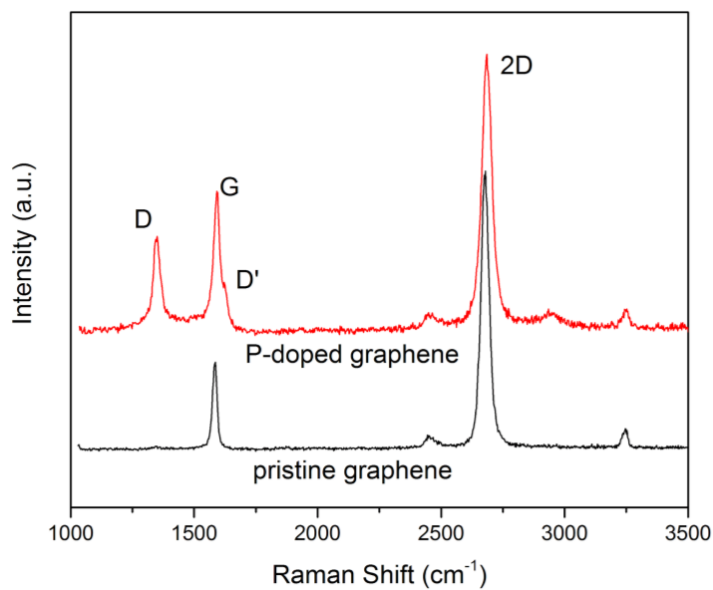

**Fig. S2**. A typical Raman spectrum of P-doped and pristine graphene. The pristine graphene was grown by using CH$_4$ as the carbon source, similar to our previous studies.

*STEM characterization*. The atomic structure of the sample is acquired by operating the aberration corrected Nion UltraSTEM 100 at the Oak Ridge National Laboratory's Center for Nanophase Materials Sciences User Facility. The sample is baked in vacuum under 160 °C for 8 hours before insertion into the microscope chamber. The electron acceleration voltage is kept at 60 kV during the operation to prevent the knock-on damage during imaging. The vacuum level at the sample volume during the experiments was kept under $3\times10^{-9}$ mbar. The final EEL spectrum is the result of adding two EEL spectra acquired during 30 s while scanning on a $5 \times 5$ Å$^2$ area containing a P atom with 4 pm/pixel and a dwell time of 16 μs/pixel. The convergence angle of electron probe is 60 mrad and the collection angle of the spectra is 96 mrad.

*STEM image simulation*. The simulation of STEM image is done by using the multi-slice method implemented in the QSTEM software package.

*First-principles calculation*. The first-principles simulation of EEL spectra uses the multiple scattering method implemented in FEFF9 with core hole approximation. The ab-initio molecular dynamics (ab-MD) is performed using density functional theory (DFT) within the general gradient approximation (GGA), in the form of Perdew-Burke-Ernzerhof's exchange-correlation functional. The time step is chosen to be 1 fs, as our calibration indicates that 1 fs time step has enough precision for predicting the dynamics (we simulated the direct exchange of P using a timestep of 0.1 fs, but found no differnce within the precision of our calculation). The lower bound is found to be 16.1 ± 0.1 eV and the upper bound is 17.3 ± 0.1 eV. All calculations were performed using the Vienna Ab-initio Simulation Package (VASP).

## 2. Overview of the P-doped graphene

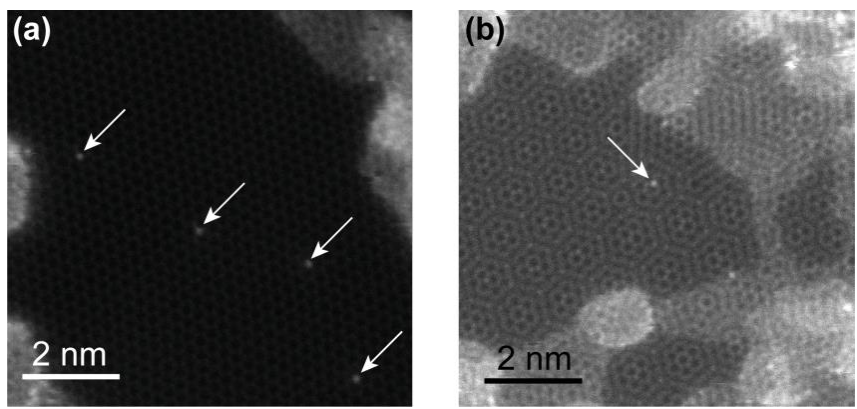

**Fig. S3**. STEM MAADF images (raw data) of the graphene area where P dopant atoms are spotted. (A) Four three-coordinated P atoms within a clean area (approx. 10 nm in size) of a single-layer graphene sheet. (B) A three-coordinated P atom embedded in a clean double-layer graphene sheet (approx. 8 nm in size). All P atoms are marked by arrows.

## 3. EELS characterization of P dopants

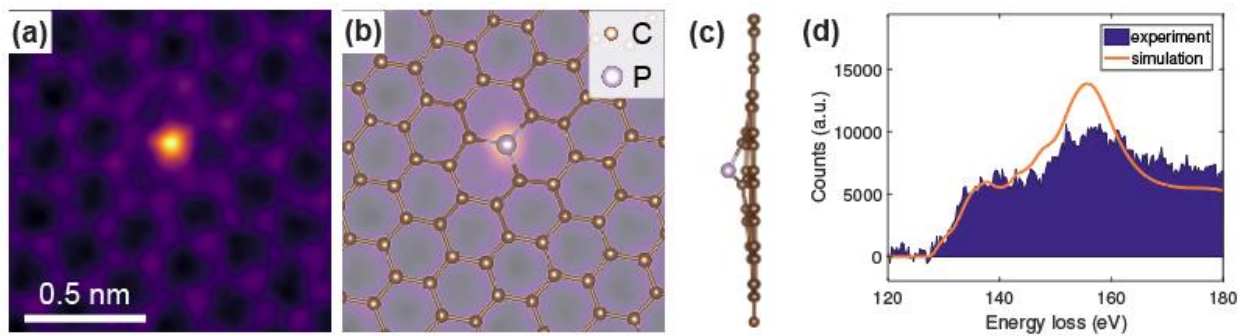

**Fig. S4.** (a) A gaussian-blurred STEM MAADF image (b) overlaid with the DFT calculated P-doped graphene structure, and (c) side view of the lowest energy configuration. (d) The EELS

acquired in experiment and multiple scattering simulation (with a core-hole approximation) of a three-coordinated P atom.

## 4. Images of Al dopant and EELS characterization

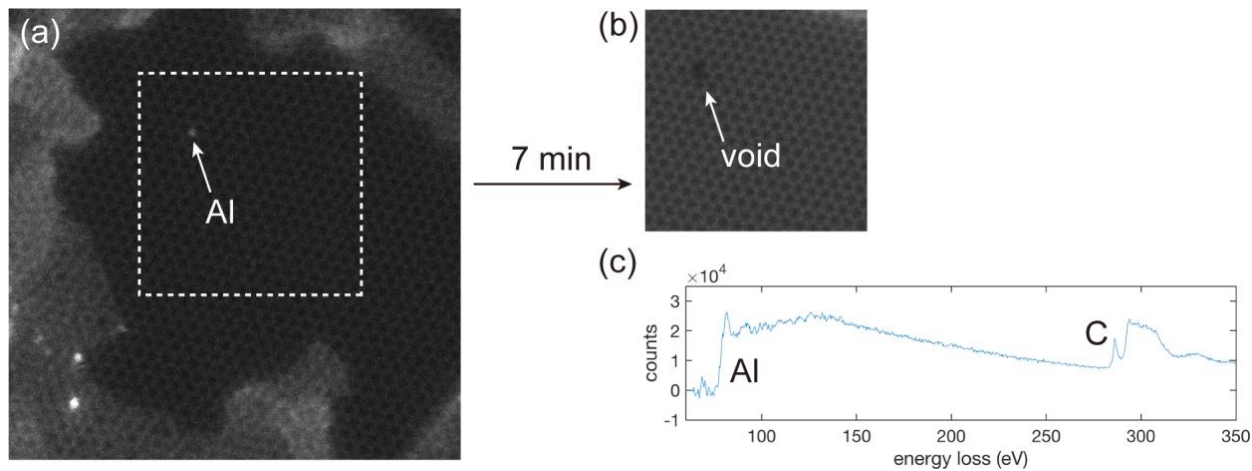

**Fig. S5**. (a) STEM image of a single layer graphene with an Al dopant marked by white arrow. (b) After 7 minutes, Al and some neighbor C atoms start to be knocked out. A void is left where Al dopant is located. A white dashed square in (a) marks the same field of view in (b). (c) The EELS showing the edges of Al and C.

## 5. Comparison of cNEB curves of various elements

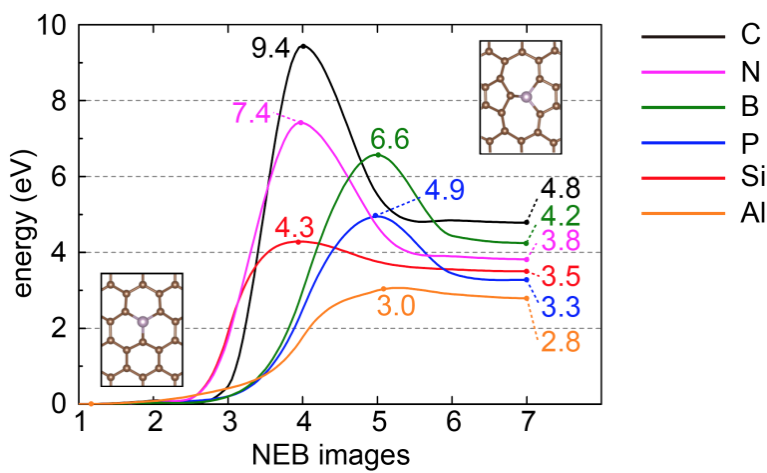

**Fig. S6**. The cNEB study of the energetic stability of the Stone-Wales transitions of different elements. The $E_a$ used in main text is obtained from subtracting the final energy from the peak energy indicated in the plot.

## 6. Comparison on replacement process of P and Si dopant

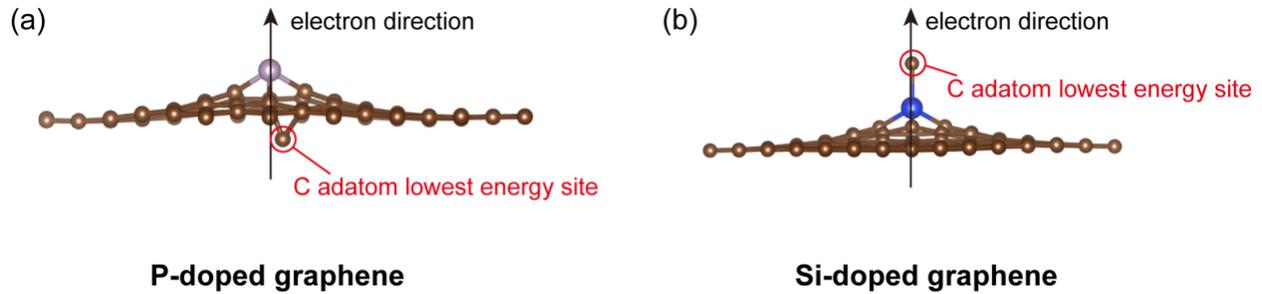

**Fig. S7**. Comparison between the C replacement process of P and Si dopant. (a) The C adatom is adsorbed at the opposite side of P dopant atom. The electron beam can transfer a proper direction of momentum to C atom to initiate the replacement process. (b) The C adatom has more probability to adsorb on top of Si dopant instead of underneath, making it impossible for C to replace Si.

## 7. P dopant in double layer graphene

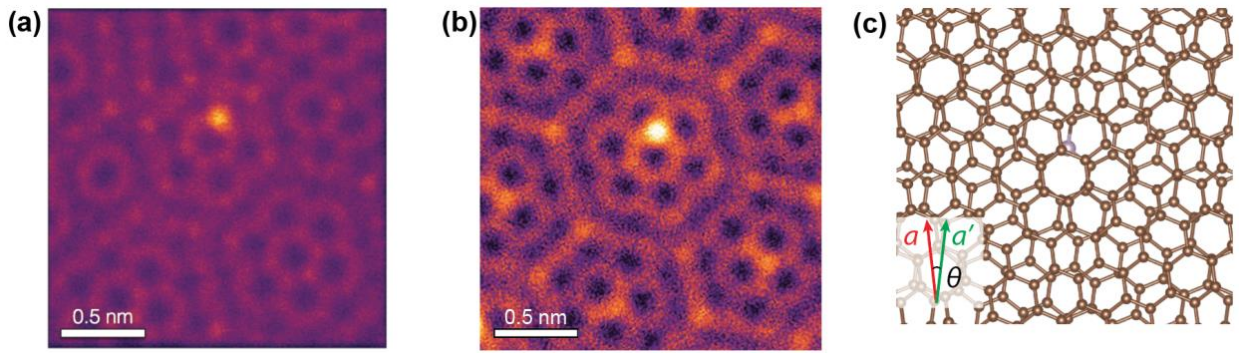

**Fig. S8.** Three-coordinated P dopant atom within a double layer graphene with **(a)** STEM and **(b)** simulated image by multislice method. **(c)** Top view and side views of the lowest energy configuration. The twist angle between two layers $\theta = 15.3°$.

## 8. Dynamics space

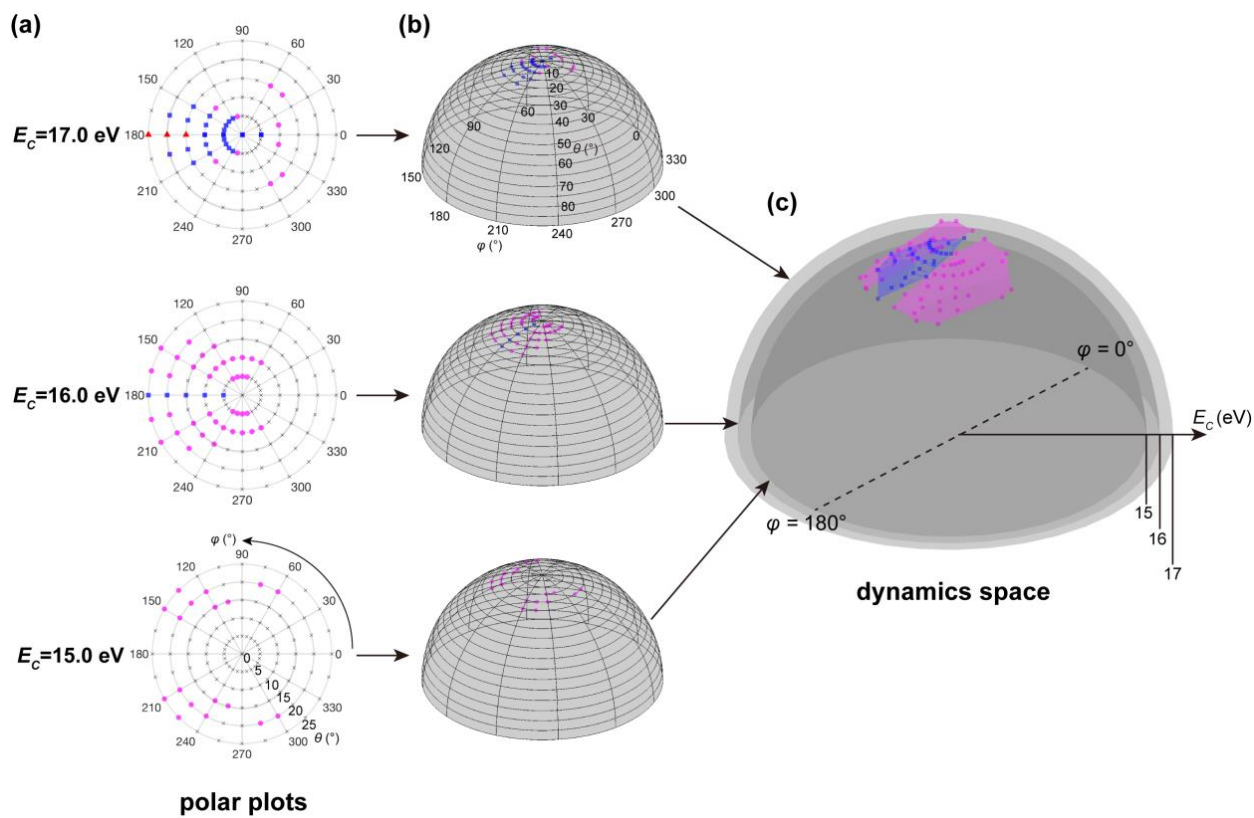

**Fig. S9.** From polar plots to dynamics space. (a) The polar plots of the distribution of dynamic processes with different C atom initial energies (15, 16, and 17 eV). (b) The dynamic processes of different energies mapped onto hemispheres, and (c) combined into the dynamics space. Only blue squares (direct exchange) and magenta circles (SW transition) are shown in the dynamics space.

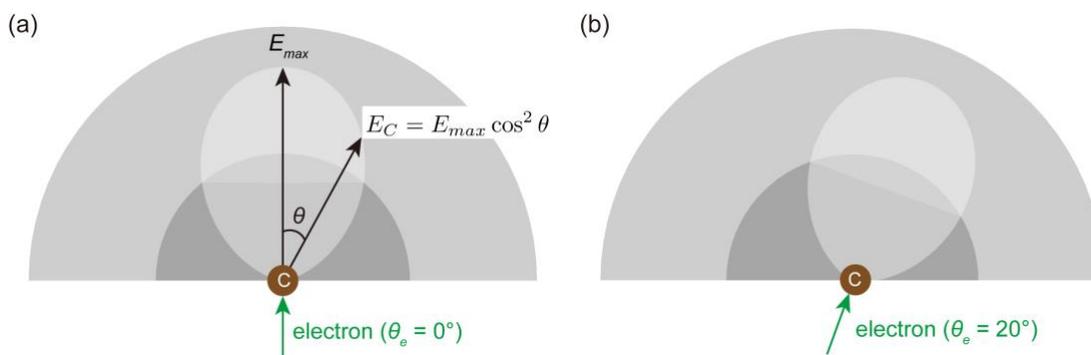

**Fig. S10.** Schematic plot of the spatial distribution of $E_C$ (dubbed as "$E_C$ ovoid" hereafter) for certain incident angles of electron: (a) $\theta_e = 0°$ and (b) $\theta_e = 20°$. The formula of $E_C$ is from [1].

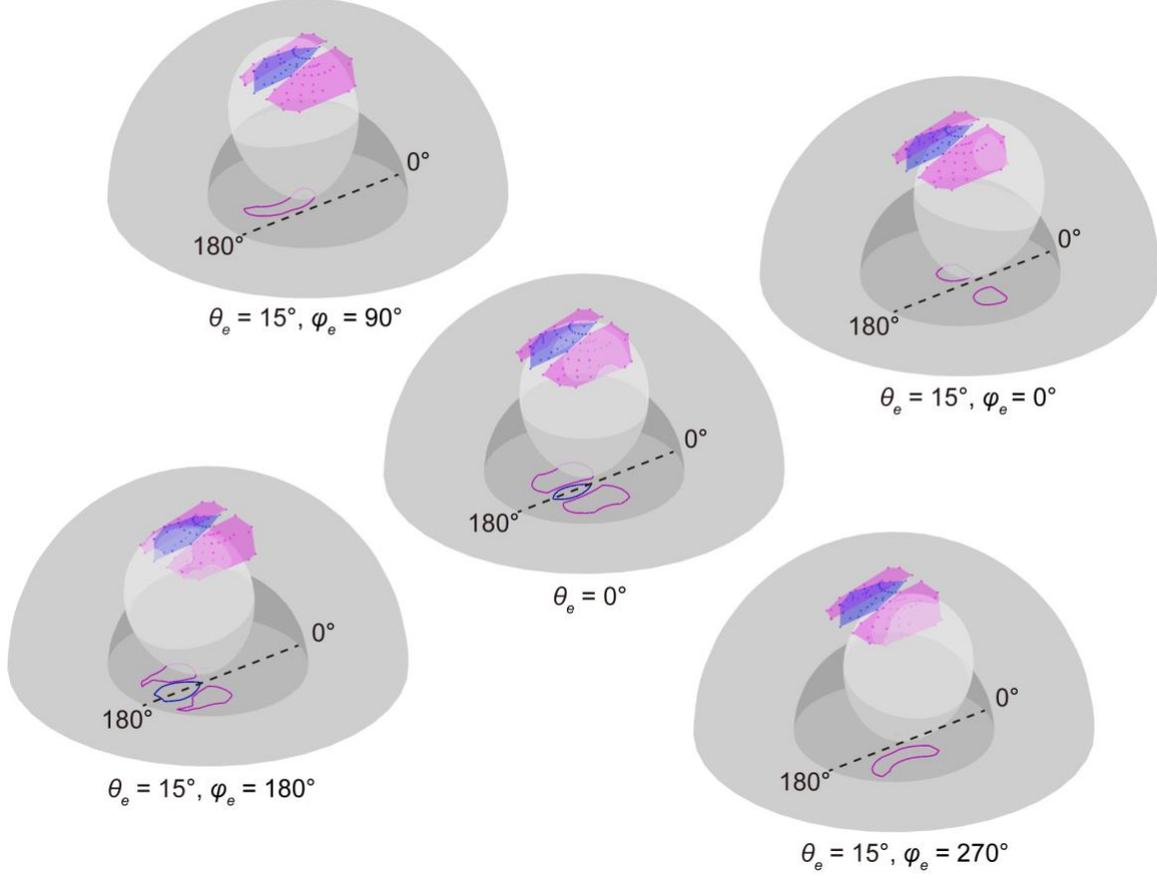

**Fig. S11.** The effective region of dynamic processes. Here, five different electron incident angles are shown ($\theta_e = 0°$, and $\varphi_e = 0°$, 90°, 180°, 270° when $\theta_e = 15°$). The intersections of the $E_C$ ovoids with different dynamic areas are projected to the bottom surface for a better view. These intersections represent the effective angles for the C atom initial momentum to achieve a certain dynamic process. By tilting the incident electron direction, different dynamic processes can be selectively initiated.

The scattering cross section from an initial configuration *i* to final configuration *k* at a certain electron incident angle ($\theta_e$ and $\varphi_e$) can be computed as

$$\sigma_{i \to k}(\theta_e, \varphi_e) = \int_A \frac{d\sigma}{d\Omega}(\theta, \varphi, \theta_e, \varphi_e) d\Omega(\theta, \varphi) \quad (1)$$

where *A* is the intersection area of the dynamic process (e.g. the area encircled by blue contour for direct exchange, or the area encircled by magenta contour for SW transition in Fig. S11), and $d\sigma/d\Omega$ is the differential cross section of the electron-carbon scattering, which depends on the incident angles of electron ($\theta_e$ and $\varphi_e$).

# 9. Atomic engineering: manipulation decision tree

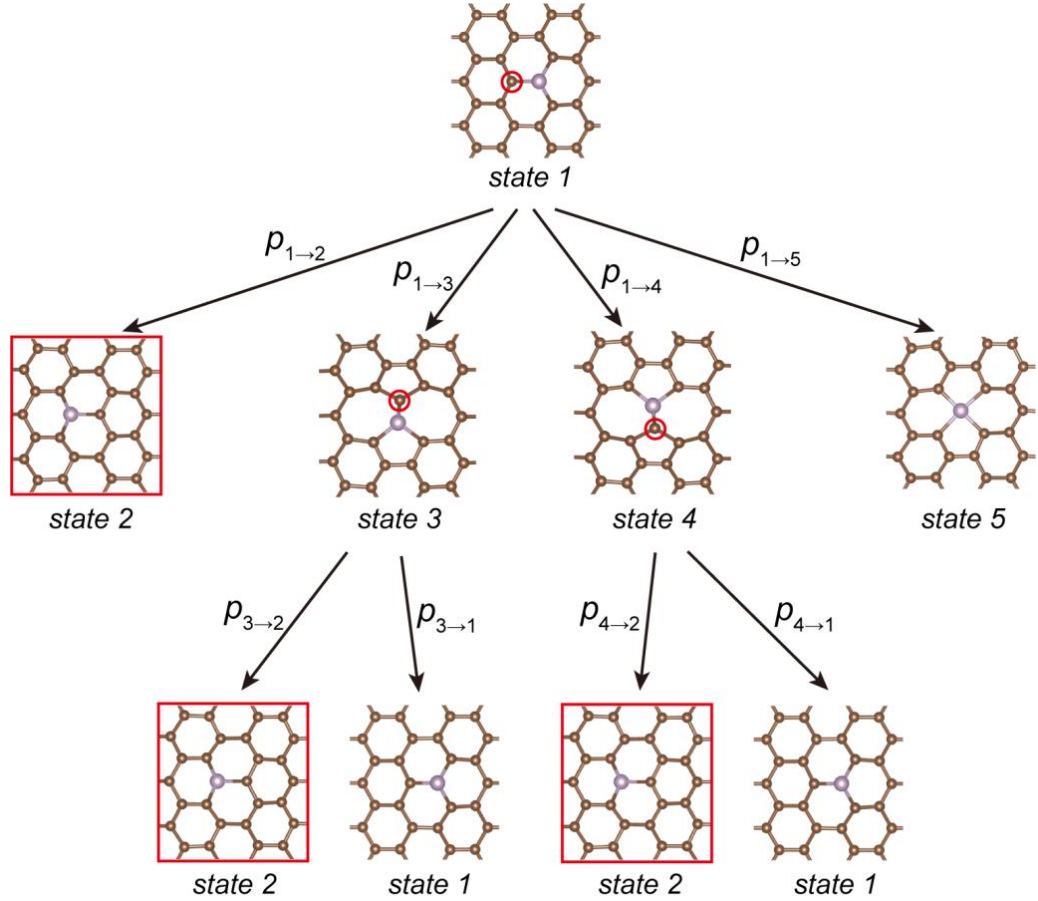

**Fig. S12.** A decision tree for engineering atom configurations in P-doped graphene. $p_{i \to k}$ stands for the probability of a dynamic process from an initial configuration *i* to final configuration *k*. We have assumed that the electron incident angles $\theta_e$ and $\varphi_e$ are fixed throughout the whole operation. The state outlined in red indicates the final desired state. Red circles indicate the target atoms of the electron irradiation.

The probability of each dynamic process can be obtained as

$$p_{i \to k}(\theta_e, \varphi_e) = \frac{\sigma_{i \to k}(\theta_e, \varphi_e)}{\sum_k \sigma_{i \to k}(\theta_e, \varphi_e)} \quad (2)$$

where $\sigma_{i \to k}(\theta_e, \varphi_e)$ can be obtained from equation (1). We can therefore maximize the probability of a specific configuration change by choosing a combination of angles that maximizes the probability of desired branches while minimizing that of undesired ones.